\documentclass[journal,twoside,web]{ieeecolor}

\usepackage{lcsys}
\usepackage{cite}
\usepackage{amsmath,amssymb,amsfonts}
\usepackage{algorithmic}
\usepackage{graphicx}
\usepackage{textcomp}
\usepackage{graphics} 
\usepackage{epsfig} 
\usepackage{mathptmx} 
\usepackage{times} 
\usepackage{booktabs}
\usepackage{csquotes}
\usepackage{multirow}
\usepackage{eso-pic}
\AddToShipoutPictureBG*{%
\AtPageUpperLeft{%
\setlength\unitlength{1in}%
\hspace*{\dimexpr0.5\paperwidth\relax}
\makebox(0,-1)[c]{
\begin{tabular}{c c}
Christopher A. Orrico \emph{et al.},
Mixed-Integer MPC Strategies for Fueling and Density Control in Fusion Tokamaks, \\
To appear in
{\em IEEE Control Systems Letters},
2023,
uploaded to PURE \today \\
\end{tabular}}}}
\AddToShipoutPictureBG*{%
\AtPageUpperLeft{%
\setlength\unitlength{1in}%
\hspace*{\dimexpr0.5\paperwidth\relax}
\makebox(0,-21.5)[c]{
\footnotesize
\begin{tabular}{c c}
© 2023 IEEE. Personal use of this material is permitted. Permission from
IEEE must be obtained for all other uses, in any \\
current or future media, including reprinting/republishing
this material for advertising or promotional purposes, creating new \\
collective works, for resale or redistribution to servers or lists,
or reuse of any copyrighted component of this work in other works.
\end{tabular}}}}

\def\BibTeX{{\rm B\kern-.05em{\sc i\kern-.025em b}\kern-.08em
    T\kern-.1667em\lower.7ex\hbox{E}\kern-.125emX}}
\begin{document}
\title{Mixed-Integer MPC Strategies for Fueling and Density Control in Fusion Tokamaks}
\author{Christopher A. Orrico, Matthijs van Berkel, \IEEEmembership{Member, IEEE}, Thomas O. S. J. Bosman, W. P. M. H. Heemels, \IEEEmembership{Fellow, IEEE}, Dinesh Krishnamoorthy, \IEEEmembership{Member, IEEE}
\thanks{This work has been carried out within the framework of the EUROfusion Consortium, funded by the European Union via the Euratom Research and Training Programme (Grant Agreement No. 101052200—EUROfusion). Views and opinions expressed are however those of the author(s) only and do not necessarily reflect those of the European Union or the European Commission. Neither the European Union nor the European Commission can be held responsible for them. This publication is part of the project \textit{Balls to the Wall} ( project no. 19695) of the research programme NWO Talent Programme VIDI, financed in part by the Dutch Research Council (NWO). DIFFER is an institute of the NWO.}
\thanks{C. A. Orrico, M. van Berkel, T. O. S. J. Bosman, and D. Krishnamoorthy are with the Dept. of Mechanical Engineering, Eindhoven University of Technology, 5600 MB Eindhoven, The Netherlands and DIFFER - Dutch Institute for Fundamental Energy Research, 5612 AJ Eindhoven, The Netherlands (e-mail: c.a.orrico@tue.nl, m.vanberkel@differ.nl, t.o.s.j.bosman@differ.nl, d.krishnamoorthy@tue.nl).}
\thanks{W. P. M. H. Heemels is with the Dept. of Mechanical Engineering, Eindhoven University of Technology, 5600 MB Eindhoven, The Netherlands (e-mail: w.p.m.h.heemels@tue.nl).}
}

\maketitle
\thispagestyle{empty}

\begin{abstract}
Model predictive control (MPC) is promising for fueling and core density feedback control in nuclear fusion tokamaks, where the primary actuators, frozen hydrogen fuel pellets fired into the plasma, are discrete. Previous density feedback control approaches have only approximated pellet injection as a continuous input due to the complexity that it introduces. In this letter, we model plasma density and pellet injection as a hybrid system and propose two MPC strategies for density control: mixed-integer (MI) MPC using a conventional mixed-integer programming (MIP) solver and MPC utilizing our novel modification of the penalty term homotopy (PTH) algorithm. By relaxing the integer requirements, the PTH algorithm transforms the MIP problem into a series of continuous optimization problems, reducing computational complexity. Our novel modification to the PTH algorithm ensures that it can handle path constraints, making it viable for constrained hybrid MPC in general. Both strategies perform well with regards to reference tracking without violating path constraints and satisfy the computation time limit for real-time control of the pellet injection system. However, the computation time of the PTH-based MPC strategy consistently outpaces the conventional MI-MPC strategy. 

\end{abstract}


\section{Introduction}
\label{sec:introduction}
\IEEEPARstart{N}{uclear} fusion is a prospective source of renewable energy that has made significant progress in recent years. Nuclear fusion power is produced when deuterium (D) and tritium (T), two hydrogen isotopes, fuse into helium. When maintained at sufficiently high density and temperature, a D-T plasma will achieve \enquote{ignition,} producing net energy from fusion \cite{Freidberg2007}. However, the plasma density and temperature necessary for ignition are extremely challenging to reach and maintain. 

\begin{figure}[!b]
\centering
\parbox{2in}{\includegraphics[width=2in]{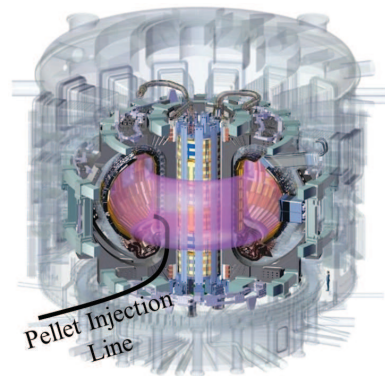}}
\caption{The pellet injection line, which fires pellets into the plasma from inner side of the plasma torus, overlaid on a cutaway diagram of the ITER tokamak. Image courtesy of the ITER Organization, with modifications \cite{ITER_CS}}
\label{fig:ITERPIS}
\end{figure}

\begin{figure}[!t]
\centering 
\parbox{\columnwidth}{\includegraphics[width=\columnwidth]{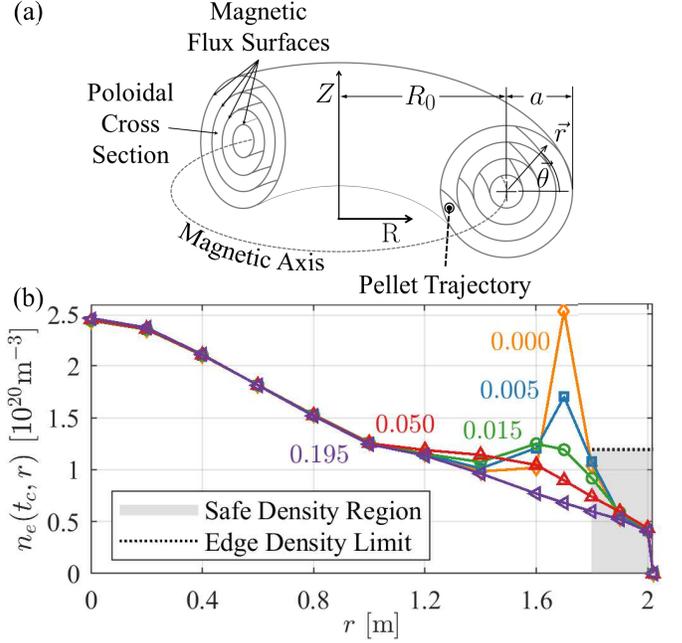}}
\caption{(a) Diagram showing nested flux surfaces of the cylindrical tokamak plasma approximation, where a fuel pellet fired along the pellet injection line in Fig. \ref{fig:ITERPIS} will ablate near the plasma edge. Here, $R_0$ is the major radius, $a$ is the minor radius, and $\theta$ is the angle from the midplane. (b) The $1\textrm{D}$ plasma response to a fuel pellet ablating at $r = 1.7$ m, for $t_c = \{0.000, 0.005, 0.015, 0.050, 0.195\}$ s after the pellet penetrates the plasma.}
\label{fig:FluxSurfToProfile}
\end{figure}

The fusion power produced by the D-T reaction scales with the square of the plasma electron density ($n_e$) \cite{Freidberg2007}. Because the torus-shaped magnetic topology confining the plasma is difficult to penetrate with particles for high density core plasma fueling, large tokamaks, such as ITER, will need to use frozen hydrogen pellets fired at a velocity sufficient to penetrate into the plasma edge for fueling (cf. Fig. \ref{fig:ITERPIS}) \cite{Baylor2007, Pegourie_2009, Baylor2016}. The particle flux from fired pellets must maintain the desired fueling rate without exceeding the safety-critical edge density limit (cf. Fig. \ref{fig:FluxSurfToProfile}b) \cite{Greenwald1988}. Violating this limit leads to a plasma disruption that can severely damage the tokamak. The control objective is to maximize the core density (and thus fusion power) without violating the edge density limit \cite{Baylor2007}. 

Plasma core density and temperature can be modeled by continuous nonlinear coupled drift-diffusion partial differential equations (PDEs) \cite{Blanken2018, Sevillano2012}, with hybrid system dynamics arising when the controller makes a binary decision to launch a pellet or not \cite{Bemporad1999}. If the controller can choose from a set of pellet frequencies, sizes, or fuel compositions, this binary set extends to a larger combinatorial set of discrete actuation decisions. Moreover, the controller cannot exceed the maximum computation time, determined by the pellet injection system's maximum repetition rate \cite{Baylor2016}. Due to the complexity that pellet fueling introduces, literature on tokamak core fueling and density control has only integrated pellet fueling as a discrete feedforward signal \cite{Garzotti2011, Vincenzi2015, Ravensbergen2017} or as a continuous actuator for feedback control \cite{Lang2018, Bosman2021, Bosman2022}. We seek to progress core density control by integerating pellet fueling as a discrete actuator. 

Hybrid model predictive control (MPC) \cite{Bemporad1999,Lazar2006} is a promising strategy for controlling systems with hybrid dynamics arising from discrete actuators \cite{Bemporad1999}. Moreover, by using MPC, we explicitly prevent the controller from violating safety-critical path constraints \cite{Bosman2021, Bosman2022}. Given a model of system dynamics, the MPC controller solves an optimal control program (OCP) subject to state and input constraints over a finite prediction horizon \cite{Rawlings2017}. If states or inputs must take integer values, the OCP becomes a mixed-integer programming (MIP) problem that may be $\mathcal{N} \! \mathcal{P}$-hard and, consequently, intractable for real-time control \cite{Kirches2011}. In this letter, we implement hybrid MPC as a strategy for the pellet fueling and density control problem in nuclear fusion tokamaks in a manner suitable for real-time control.

To this end, we model particle transport in a fusion plasma with discrete fuel pellet injection as a hybrid dynamical system. We then propose two real-time MI-MPC setups to solve the density control problem. The conventional strategy for solving an MI-MPC problem is to use a mixed-integer solver. Here, we compute a mixed-integer quadratic programming (MIQP) problem at each discrete-time step using the branch and bound (B\&B) algorithm of the \texttt{Gurobi} solver. Next, we develop a novel version of the penalty term homotopy (PTH) algorithm and apply it to the MI-MPC problem, such that it is solvable using a quadratic programming (QP) solver (such as \texttt{qpOASES}), rather than an MIP solver. The PTH algorithm \cite{Sager2006} transforms a MI-MPC problem into a series of continuous optimization problems by relaxing the integer requirements, instead penalizing non-integer solutions in the objective function \cite{Sager2006}. However, as originally designed \cite{Sager2006}, the PTH algorithm may not be able to satisfy both integer requirements and path constraints. By adding a logarithmic barrier term, we modify the PTH algorithm in a novel way to be viable for path-constrained hybrid MPC problems. We then compare the performance of the two strategies in terms of their reference tracking error and computation time.

\section{Problem Formulation}
\label{sec:problemFormulation}

During a tokamak plasma discharge, we plan to maintain high density in the core of the plasma using hydrogen fuel pellets \cite{Baylor2016}. In Section \ref{subsec:driftDiffusionModel}, we model the tokamak plasma core temperature and density using a system of coupled $1\textrm{D}$ drift-diffusion equations, integrating discrete pellet actuation as a particle source. We then translate this model into a form appropriate for control. In Section \ref{subsec:ControlProblem}, we formulate the control problem, introducing key parameters and constraints.

\subsection{Drift-Diffusion Model}
\label{subsec:driftDiffusionModel}
In MPC, the model of the system dynamics is crucial for predicting the state trajectory. Fusion plasma transport dynamics across nested flux surfaces (cf. Fig. \ref{fig:FluxSurfToProfile}a) are modeled as a system of continuous coupled nonlinear PDEs \cite{Freidberg2007}. As computing the evolution of these equations in $3\textrm{D}$ space is impractical for real-time control, we approximate transport in a toroidal plasma with a simplified $1\textrm{D}$ drift-diffusion model by adopting several realistic assumptions. First, we assume that the plasma particles (electrons and ions) travel much faster along the nested flux surfaces in Fig. \ref{fig:FluxSurfToProfile}a than they travel perpendicular to the flux surfaces \cite{Freidberg2007}. Consequently, the density and temperature are constant along the flux surfaces, yielding a $1\textrm{D}$ model of the density and temperature gradients normal to said surfaces. Next, we approximate the plasma as a cylindrical torus, further simplifying the density and temperature gradients as functions of the radius $r$ (cf. Fig. \ref{fig:FluxSurfToProfile}a). The $1\textrm{D}$ coupled nonlinear PDEs between the plasma density and temperature are written as

\begin{subequations}
\begin{align}
    \begin{split}
	\frac{\partial n_e}{\partial t_c} &= S_{n_e}(r,t_c) + \frac{1}{r} \biggl\{ \frac{\partial}{\partial r} rD_n \biggl[ \frac{\partial n_e}{\partial r} + \frac{n_e}{T_e}\frac{\partial T_e}{\partial r} +\\
    &\; + \frac{D_B}{4\eta_\parallel}\frac{B_p}{T_e r}\left(B_p+r\frac{\partial B_p}{\partial r}\right) \biggr] + \nu \frac{\partial}{\partial r}\left(rn_e\right) \biggr\} ,\label{eq:dndt}
    \end{split}\\
	\begin{split}
		\frac{\partial T_e}{\partial t_c} &= \frac{1}{3n_e} \biggl[ \frac{1}{r}\frac{\partial}{\partial r} \left(rn_e\chi \frac{\partial T_e}{\partial r} \right) + S_{T_e}(r,t_c)\biggr], \label{eq:dTdt}
	\end{split}
\end{align} \label{eq:PDE}
\end{subequations} 

\noindent where $n_e$ is the electron density, $T_e$ is the electron temperature, $B_p$ is the poloidal magnetic field, $r$ is the radius from the plasma core, $t_c$ is continuous time, $\eta_{\parallel}$ is the Spitzer resistivity, and the magnetic diffusion coefficient is $D_B=\eta_{\parallel}/\mu_0$  \cite{Freidberg2007, Blanken2018}. The device parameters (including minor radius $a$) are set to those of the ITER tokamak \cite{Garzotti2019}. The particle diffusion, the particle drift, and the thermal diffusion coefficients are $D_n$, $\nu$, and $\chi$, respectively \cite{Na2019, Polevoi2002}. Here, a fired fuel pellet is a density source $S_{n_e}(r,t_c)$ at the radius of deposition over the time period of ablation $\mathcal{O}(1)$ ms \cite{Pegourie_2009}. $S_{T_e}(r,t_c)$ is the plasma heat source from fusion reactions and auxiliary heating, which we assume to be constant in $r$ and $t_c$. For a detailed discussion on modeling fusion plasma transport, the reader is referred to \cite{Freidberg2007}. While \eqref{eq:PDE} simplifies fusion plasma dynamics, it is straightforward to analyze. Consequently, \eqref{eq:PDE} is ideal as a proof-of-concept model against which to test control strategies.

Discretizing \eqref{eq:PDE} with respect to $r$, we construct a piecewise smooth ODE for $n_e$ and $T_e$ using the first-order finite difference method. Boundary conditions are set as $n_e|_{r=a + \lambda}=0$, $T_e|_{r=a + \lambda}=0$, $\nabla n_e|_{r=0}=0$, and $\nabla T_e|_{r=0}=0$. Here the edge is a perfect sink at the virtual domain node $r = a + \lambda$ \cite{Blanken2018}, where the minor radius $a = 2$ m and $\lambda$ is the "scrape-off layer" width \cite{Garzotti2019}. 

To be consistent with control conventions, we translate \eqref{eq:PDE} and nuclear fusion specific terminology to a state-space model, constraints, and a state reference. We begin by defining the continuous time ($t_c$) state space description of the dynamics as

\begin{subequations}
\begin{align}
     \dot{x}(t_c) &= f\left(x(t_c), u(t_c) \right), \label{eq:xdot} \\
    \begin{split}
    x(t_c) &= [n_e(t_c,r_1) ~ n_{e}(t_c,r_2) ~ \ldots ~n_e(t_c,r_{n/2}), \\ &\; \quad ~T_e(t_c,r_1) ~ T_e(t_c,r_2) ~\ldots~ T_e(t_c,r_{n/2}]^{\top}, \label{eq:x} 
    \end{split}\\
    \begin{split}
    u(t_c) &= [u_1(t_c) ~u_2(t_c) ~\ldots~ u_m(t_c)]^{\top}. \label{eq:u}  
    \end{split}
\end{align} \label{eq:statedynamics}  
\end{subequations}

\noindent The nonlinear, piecewise smooth ODE discretized in $r$ from \eqref{eq:PDE} is given as the vector field $f: \mathbb{R}^{n} \times \mathbb{R}^{m} \rightarrow \mathbb{R}^{n}$, and is a function of the state vector $x(t_c) \in \mathbb{R}^{n}$ for $n/2$ discrete radial nodes $[r_1, r_2, \ldots, r_{n/2}]$ and input vector $u \in \mathbb{R}^{m}$ for $m$ pellet launchers. We take $n = 26$ corresponding to $13$ radial nodes $r = \{0, 0.2, 0.4,...,1.6, 1.7,..., 2.0\}$ m. For simplicity, we take $m=1$ for a single pellet launcher depositing pellets at $r = 1.7$ m (cf. Fig. \ref{fig:FluxSurfToProfile}b). The pellets have a fixed size of $6\times 10^{21}$ D atoms, which is the fuel pellet size planned for ITER \cite{Baylor2016}. 

To simulate the plasma behavior during the \enquote{flat-top} of an ITER $15$ MA plasma discharge \cite{Garzotti2019}, we discretize the hybrid system dynamics \eqref{eq:statedynamics} in time using the Runge-Kutta 4th order (RK4) method with a sampling time of $5$ ms, which is sufficiently fast to ensure numerical stability and is consistent with the rate of pellet ablation in the plasma \cite{Franklin1997, Pegourie_2009}. The controller supplies an input $u(t)$, computed by the controllers designed in Section \ref{sec:MPCapproach}, in feedback to the simulation.

\subsection{Control Problem Description}
\label{subsec:ControlProblem}

The objective of pellet fueling and density control is to track a reference density without violating safety-critical path constraints. For instance, because fusion power scales $\propto n_e^2$, the reference may seek to maximize the core density. In this section, we outline the general control problem to be solved in Section \ref{sec:MPCapproach}.

For the MPC prediction model, which is distinct from the non-linear plasma dynamics simulation above, we take the Jacobian linearization of \eqref{eq:statedynamics} around a desired local equilibrium point \cite{Franklin1997, Rawlings2017}. We compute the exact discretization of the Jacobian linearization using the zero-order hold (ZOH) method \cite{Franklin1997}, selecting $T_{s,\textrm{ZOH}} = 5$ ms both for consistency with the simulated system RK4 time discretization and that a fired pellet ablating for $5$ ms results in a constant density source over $T_{s,\textrm{ZOH}}$. Thus, the linearization approximates \eqref{eq:PDE} as a linear time-invariant (LTI) state space model with state and input matrices $A$ and $B$ in discrete time $t$.

Finally, the maximum repetition rate of the pellet injection system must be built into the prediction model. We base our model off of the ITER tokamak pellet injection system, which has 2 pellet injection lines operating with a maximum repetition rate of $4$ Hz per injector \cite{Baylor2016}. As the system requires a control decision every $\approx 100$ ms, we set the control sampling time $T_s = 100$ ms and define discrete time $t \in \mathbb{N}$ with respect to continuous time as $t_c = tT_s$. The prediction model now must include the first $5$ ms of zero-order hold actuation and subsequent $95$ ms of state evolution with $u_{k|t} = 0$ to account for the time between decisions. Accordingly, we extend the LTI state prediction model as

\begin{equation}
    x(t+1) = \bar{A}x(t)+\bar{B}u(t), \label{eq:ZOHfix}
\end{equation}

\noindent where $\bar{A} = A^{\tau}$ and $\bar{B} = A^{\tau-1}B$, defining $\tau = T_s/T_{s,\textrm{ZOH}} \in \mathbb{N}$, such that the prediction model now incorporates one control decision every $T_s = 100$ ms \cite{Rawlings2017}. 

The edge density limit of a fusion plasma, given as $n_{Gw} = I_p/(\pi a^2)$ (where $I_p$ [MA] is the total plasma current) \cite{Greenwald1988}, places hard path constraints on the controller (cf. Fig \ref{fig:FluxSurfToProfile}b) and is central to the controller design. We construct an element-wise inequality with matrix $G$ and vector $h$ that MPC stricly enforces $n_e(t,r=1.8\textrm{ m}) \leq n_{Gw}$ as 
\begin{equation}
Gx - h \leq 0. \label{eq:constraintsdef}
\end{equation}

For reference tracking, the line-averaged electron density $\bar{n}_e(t)$ \cite{Heald1965} is the key diagnostic measurement, calculated as 

\begin{equation}
 \bar{n}_e = \frac{1}{a}\int_0^an_edr \!~ . \label{eq:lineAvgDens}
\end{equation}

\noindent We set the state reference $x^r(t) = \bar{n}_e^r(t)$ for a desired core density \cite{Bosman2021kalman}. The control objective is now to design a controller capable of tracking $x^r(t)$ without violating constraints \eqref{eq:constraintsdef} in real-time (given as the computation time limit $T_s$). 

\section{Model Predictive Control Strategies}
\label{sec:MPCapproach}
To solve the problem formulated in the previous section, we propose to use MPC. This method solves a receding horizon optimization problem based on the measured state online and applies the first control action to the plant, repeating this process for each discrete-time step \cite{Rawlings2017}. The optimization problem seeks to minimize an objective function with respect to the optimization variable $u_{k|t}$. The objective function (see \eqref{eq:objective}, \eqref{eq:PTHobjective} below) consists of stage and terminal costs, which are functions of the current state $x(t)$, the inputs $u_{k|t}$, and the predicted state trajectory $x_{k|t}$ for $k = \{0, 1,\ldots,N\}$ discrete-time steps after $t$. Solving the resulting receding horizon optimization problem yields the optimal $u^*_{k|t}$. 

\subsection{Mixed-Integer MPC}
\label{subsec:MIQP}
MI-MPC is a special case of MPC, where the set of admissible states or inputs must take integer values \cite{Kirches2011}. In the case of pellet injection control, we enforce a binary integer constraint \eqref{eq:binary} on all predicted inputs $u_{k|t}$. At $k|t$, no pellet is fired if $u_{k|t} = 0$, whereas a pellet is fired if $u_{k|t} = 1$. The MI-MPC problem is defined at $t$ for the current state $x(t)$ as

\begin{subequations}
\begin{align}
\min_{u_{k|t}} \|(x_{N|t}-x^r(t))\|^2_P ~+& \sum_{k=0}^{N-1} \|(x_{k|t}-x^r(t))\|^2_Q + \|u_{k|t}\|^2_R  \label{eq:objective} \\ 
\textrm{s.t.}&  \nonumber \\
x_{k+1|t} = \bar{A}x(t)+\bar{B}u(t),  \quad & k \in \{0,1,\ldots,N-1\} \label{eq:stateEvo} \\
Gx_{k|t}-h \leq 0, \quad  & k \in \{0,1,\ldots,N\} \label{eq:lincons}    \\
u_{k|t} \in \{0,1\}, \quad  &  k \in \{0,1,\ldots,N-1\} \label{eq:binary} \\
x_{0|t} = x(t). \quad &
\end{align} \label{eq:MIQP}
\end{subequations} 

\noindent Above, the norm $\|a\|^2_B := a^\top B a$ yields the convex objective function \eqref{eq:objective}, wherein $Q \in \mathbb{R}^{n\times n}$, $R \in \mathbb{R}^{m\times m}$, and $P \in \mathbb{R}^{n\times n}$ are diagonal tuning matrices that penalize the state error, the input cost, and the terminal state error, respectively. Q and P scale the density state to 1 and are weighted to account for non-uniform radial discretization, while the temperature states are weighted to 0. Lastly, we let $R=0$ as we are not concerned with penalizing control actuation. We implement the MI-MPC problem using \texttt{CasADi} and the \texttt{Yalmip} toolbox \cite{Andersson2019, Lofberg2004}. We warm-start $u_{k|t}$ with $u^*_{k|t-1}$. The B\&B algorithm (computed online with the MIQP solver \texttt{Gurobi}) searches a tree of possible candidates for $u^*_{k|t}$ to find the optimal control sequence \cite{gurobi}. For further details on the B\&B algorithm, we direct the reader to \cite{gurobi}. 

\begin{table}[b!]
\centering
\begin{tabular}{rl}
\toprule[2pt]
\multicolumn{2}{l}{\textbf{Algorithm 1} Modified Penalty Term Homotopy} \\ \midrule
1: & Set $j\leftarrow0$, $\gamma_j \leftarrow 0$, and $\beta_j \leftarrow 0$\\
2: & \textbf{while} $\left( u_{k|t} > \epsilon^\dagger ~ \textrm{\textbf{or}} ~ 1 - u_{k|t} > \epsilon^\dagger\right) ~\forall k \in \{0,\ldots,N-1\} ~\ldots$ \\
& \hspace{6mm} \textbf{and} $j < j_{\textrm{max}}{}^\dagger$ \textbf{do:} \\
3: & \hspace{3mm} Solve the relaxed optimization problem \eqref{eq:PTHMPC}\\
4: & \hspace{3mm} Set $j \leftarrow j+1$ \\
5: & \hspace{3mm} Set $\beta_j \leftarrow \beta_{\textrm{init}} \beta_{\textrm{inc}}^{j-1}$, $\gamma_j \leftarrow \gamma_{\textrm{init}} \gamma_{\textrm{inc}}^{j-1} {}^\ddagger$   \\
6: & \textbf{end while} \\
\bottomrule
\multicolumn{2}{p{230pt}}{$^\dagger \!\epsilon$, a user-defined tolerance, is sufficiently small such that the rounding error on $u_{k|t}$ will not result in violation of state constraints and $j_{max}$ guarantees that the algorithm is finite. If $j = j_{max}$, Algorithm 1 terminates and sets $u_{k|t} \leftarrow 0$. Sager provides additional stopping criteria to ensure sufficiently refined control discretization \cite{Sager2006}, which we omit here (see Section \ref{subsec:PTHmod}).}\\
\multicolumn{2}{p{230pt}}{$^\ddagger$Setting $\gamma_{\textrm{init}} = 0$ results in the unmodified PTH algorithm.}
\end{tabular}
\label{alg:PTH}
\end{table}

\subsection{Penalty Term Homotopy MPC}
\label{subsec:PTHQP}
When the B\&B algorithm solves \eqref{eq:MIQP} for the optimal control sequence, the size of the search tree (and thus the number of QPs the MIQP solver computes) increases exponentially with $N$. Sager \cite{Sager2006} proposes the PTH algorithm as a means to compute the optimal solution to a MIQP problem using a QP solver rather than a MIQP solver. By relaxing \eqref{eq:binary} to $u_{k|t} \in [0,1]$, \eqref{eq:objective} transforms into a continuous function of $u_{k|t}$ in \eqref{eq:PTHobjective}. To implicitly enforce integer requirements, the concave penalty term $u_{k|t}\beta_j(1-u_{k|t})$ in \eqref{eq:PTHobjective} replaces the binary constraint on $u_{k|t}$ in \eqref{eq:binary}. Each iteration of Algorithm 1 solves \eqref{eq:PTHMPC}, resulting in a series of QPs. Consequently, the computation time scales $\mathcal{O}(2N)$ rather than $\mathcal{O}(2^N)$ (note that the value 2 here indicates a binary decision variable).

At discrete-time step $t$, \texttt{qpOASES} \cite{Ferreau2014} solves the warm-started QP for each iteration $j$ of Algorithm 1,

\begin{subequations}
\begin{align}  
\begin{split}
\min_{u_{k|t}} \|(x_{N|t}-x^r(t))\|^2_P ~ - ~ & \gamma_j \ln \left(h-Gx_{N|t} \right) ~+ \\
+ \sum_{k=0}^{N-1}\biggl[ \|(x_{k|t}-x^r(t))\|^2_Q ~+~ & \|u_{k|t}\|^2_R ~+  \\
+~ u_{k|t} \beta_j (1-u_{k|t})^{\top} -~ & \gamma_j \ln \left(h-Gx_{k|t} \right)  \biggr], \label{eq:PTHobjective} 
\end{split}  \\
\textrm{s.t.} & \nonumber \\
x_{k+1|t} = \bar{A}x(t)+\bar{B}u(t), \quad & k \in \{0,1,\ldots,N-1\} \label{eq:PTHstateEvo} \\
Gx_{k|t}-h \leq 0, \quad & k \in \{0,1,\ldots,N\} \label{eq:PTHlincons}\\
u_{k|t} \in [0,1], \quad & k \in \{0,1,\ldots,N-1\} \label{eq:nonbinary} \\ 
x_{0|t} = x(t), \quad & 
\end{align} \label{eq:PTHMPC}
\end{subequations}

\noindent in which $\beta_j = \beta_{\textrm{init}} \beta_{\textrm{inc}}^{j-1}$ and increases for each $j$ until $u_{k|t} \in \{0,1\}$ for all $ k \in \{0,1,\ldots,N-1\}$ (or another stopping criterion is met) \cite{Sager2006}. We tune $\beta_{init}$ and $\beta_{inc}$ for performance, starting with the initial values that \cite{Sager2006} proposes of $\beta_{init} = 1\times 10^{-4}$ and $\beta_{inc} = 2$ for systems with variables scaled to 1. Increasing $\beta_{init}$ and $\beta_{inc}$ reduces the average algorithm iterations at each time step at the cost of reference-tracking performance. Note that in the original PTH algorithm, $\gamma_j=0$. The logarithmic barrier term $-\gamma_j \ln (h-Gx_{k|t})$ is a novel addition to the PTH algorithm and will be addressed in Section \ref{subsec:PTHmod}.

\subsection{Penalty Term Homotopy MPC with Constraints}
\label{subsec:PTHmod}

\begin{table}[b!]
\caption{PTH-MPC Sticking Behavior}
\centering
\begin{tabular}{rllllll}
\toprule[1pt]
$ j $ & $\beta_j$  & $u^*_{0|0}$ & $u^*_{1|0}$ & $u^*_{2|0}$ & $u^*_{3|0}$ & $u^*_{4|0}$ \\  \midrule
 0 & 0 & 0.9364 & 0.5382 & 0 & 0.2498 & 0.3086 \\
 $1$ & $0.1$ & 0.9364 & 0.9239 & 0 & 0 & 0 \\
 $\vdots$ & $\vdots$ & $\vdots$ & $\vdots$ & $\vdots$ &  $\vdots$ & $\vdots$ \\
 $j_{\textrm{max}}$ & $\beta_{j_\textrm{max}}$ & 0.9364 & 0.9239 & 0 & 0 & 0 \\
\bottomrule
\end{tabular}
\label{tab:sticky}
\end{table}

\begin{figure*}[t!]
\centering
\parbox{\textwidth}{\includegraphics[width=\textwidth]{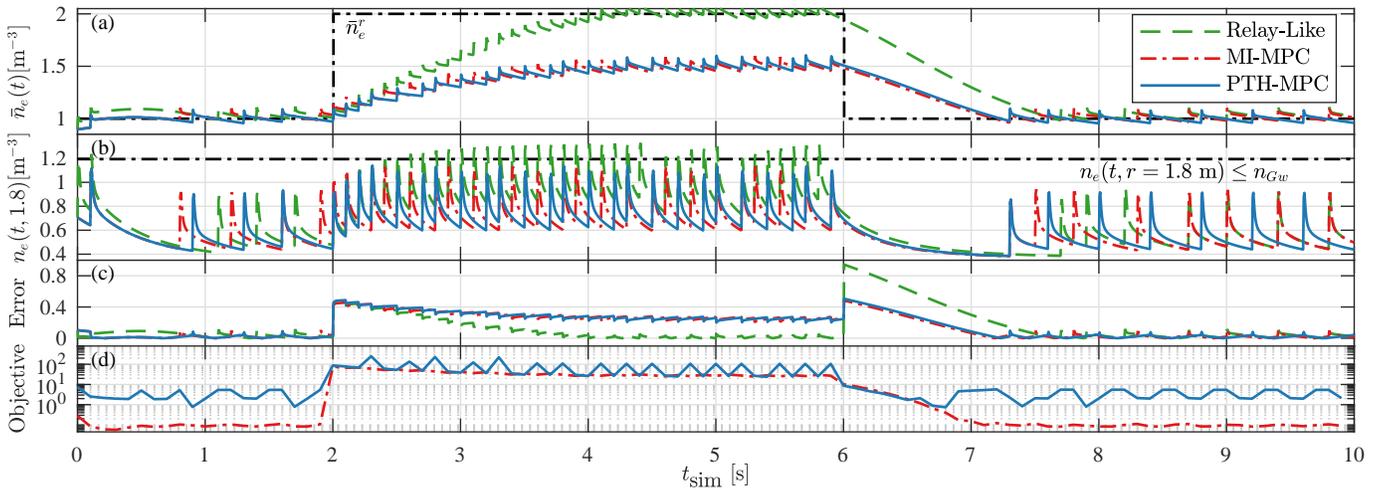}}
\caption{(a) The line-averaged density $\bar{n}_e(t)[\textrm{m}^{-3}]$ (calculated from $x(t)$ using \eqref{eq:lineAvgDens}) for $N=5$. The grey dashed line indicates $x^r(t) = \bar{n}_e^r(t)$. (b) The edge density $n_e(t, r=1.8\textrm{ m})[\textrm{m}^{-3}]$ with the density limit \eqref{eq:constraintsdef} indicated as a black dashed line. Each pellet actuation is visible as a jump in density. (c) The reference tracking error of $\bar{n}_{e}(t)$ relative to $\bar{n}_e^r(t)$. (d) The value of the objective function of each MPC strategy.}
\label{fig:ctrlTrajectory}
\end{figure*}

In its original form \cite{Sager2006}, the PTH algorithm is not guaranteed to produce feasible solutions for systems with path constraints. By relaxing the integer requirement, the optimal solution $u^*_{k|t}$, obtained by minimizing \eqref{eq:PTHobjective} in the presence of path constraints \eqref{eq:PTHlincons}, may not converge to the same solution as the antecedent MIQP \eqref{eq:MIQP}. When the descent direction of \eqref{eq:PTHobjective} with the concave $u_{k|t}\beta_j(1-u_{k|t})$ is towards path constraints, \eqref{eq:PTHMPC} may get "stuck" on an active path constraint with $u^*_{k|t} \notin \{0,1\}$, even as $\beta_j\rightarrow \infty$. Table \ref{tab:sticky} demonstrates this behavior for the control trajectory in Fig. \ref{fig:ctrlTrajectory}.

To resolve this issue, we add the logarithmic barrier term $-\gamma_j \ln (h-Gx_{k|t})$ to \eqref{eq:PTHobjective} (with $\gamma_j = \gamma_{\textrm{init}} \gamma_{\textrm{inc}}^{j-1}$ defined in Algorithm 1). As $\gamma_j$ increases,  the logarithmic barrier term drives the QP solver solution for $u^*_{k|t}$ away from the path constraints and towards the MIQP solution. To the best of our knowledge, this modification of the PTH algorithm, which makes it broadly viable for path-constrained MI-MPC problems, is novel. For the various $N$ values in Section \ref{sec:results}, we tune $\gamma_j$ relative to $\beta_j$ as $\beta_{init} = 0.5/N$, $\beta_{inc} = 3$, and $\gamma_j=\beta_{j+1}$ such that $j<j_{max} \forall t$.

\section{Results}
\label{sec:results}
The MPC strategies are evaluated for real-time performance with two metrics: the reference tracking error and the computation time. The controllers, tested with prediction horizons $N \in \{2,5,10,20\}$ in a feedback loop, are also compared to a relay-like control law: $\{u(t) \leftarrow 1 \textrm{\textbf{ if }} \bar{n}_e(t) < \bar{n}_e^r(t) \textrm{\textbf{, else }} u(t) \leftarrow 0 \}$. The RK4 discretization of \eqref{eq:PDE} simulates the nonlinear system dynamics over a period $t_{\textrm{sim}} = [0,10]$ s. 

\subsection{Reference Tracking}
\label{subsec:refTracking}

\begin{table}[b!]
\caption{Mean Reference Tracking Error}
\centering
\begin{tabular}{ccccc}
\toprule
$N$& \multicolumn{2}{c}{Reachable State $\%$ Error} & \multicolumn{2}{c}{Unreachable State $\%$ Error} \\
\cmidrule(lr){2-3}\cmidrule(lr){4-5}
& MI-MPC & PTH-MPC & MI-MPC & PTH-MPC\\ 
\midrule
2 & 2.34 & 2.13 & 29.4 & 30.2 \\
5 & 2.34 & 2.09 & 29.4 & 29.7 \\
10 & 2.34 & 2.12 & 29.4 & 30.3 \\
20 & 2.34 & 2.11 & 29.4 & 29.8 \\ 
\bottomrule \\
\multicolumn{5}{p{230pt}}{The arithmetic mean error between $\bar{n}_e(t)$ relative to $\bar{n}_e^r(t)$ is calculated over $t_{\textrm{sim}} = 0,...,2,6,...,10$ s and $t_{\textrm{sim}} = 2,...,6$ s for the reachable $\bar{n}_e^r(t)$ and unreachable $\bar{n}_e^r(t)$, respectively.}
\end{tabular}
\label{tab:error}
\end{table}

The state reference $x^r(t)$ is initially set at $\bar{n}_e(t) = 10^{20}$ m$^{-3}$, corresponding to the \enquote{flat-top} of a D-T H-mode $15$ MA ITER plasma \cite{Garzotti2019}. At $t_{\textrm{sim}}=2$ s,  $x^r(t)$ steps up to $\bar{n}_e(t) = 2\times 10^{20}$m$^{-3}$. This reference is unreachable due to the edge density limit, forcing the MPC strategies to demonstrate their ability to drive the state towards an optimal trajectory without violating safety-critical path constraints, which is desirable for exploring maximum achievable plasma density \cite{Lang2018}.

Fig. \ref{fig:ctrlTrajectory} compares the state trajectories for a prediction horizon of $N=5$. Both the MPC strategies show good reference tracking when the state reference is reachable (cf. Fig. \ref{fig:ctrlTrajectory}a) and, unlike the relay-like controller, do not violate path constraints (cf. Fig. \ref{fig:ctrlTrajectory}b). The two MPC strategies compute slightly different control policies for the same prediction horizon (cf. Fig. \ref{fig:ctrlTrajectory}a). This is a consequence of the differing objective costs (cf. Fig. \ref{fig:ctrlTrajectory}d) that arise from the addition of the penalty terms in the PTH-MPC formulation. The controller performance in Fig. \ref{fig:ctrlTrajectory} is consistent across all tested values of $N$. Mean tracking error in Table \ref{tab:error} is nearly equal for both control approaches. Furthermore, increasing $N$ has little impact on mean tracking error as the plasma transport is so rapid that the density profile settles within $\approx 200$ ms of pellet injection (cf. Fig. \ref{fig:FluxSurfToProfile}b). Consequently, a prediction horizon as short as $N=2$ can largely capture predicted plasma response to the next control decision. 

\subsection{Computation Time}
\label{subsec:compTime}

For the purposes of real-time feedback control, we require the computation time of the controllers to not exceed $100$ ms \cite{Baylor2016}. Table \ref{tab:compTime} lists the maximum computation times during the simulation in Fig. \ref{fig:ctrlTrajectory}. For all $N$ tested, the PTH-MPC control approach outperformed the MI-MPC method. The exponential growth of the MI-MPC computation time for increasing $N$ is evident for $N = 20$, jumping an order of magnitude over the computation time of the PTH-MPC solution for the same $N$. For short $N$, the PTH-MPC method is sufficiently fast to not only accommodate the ITER pellet injection system, but also the up to $80$ Hz repetition rates of the faster centrifugal pellet launching systems \cite{Lang2019}. 

\begin{figure}[t!]
\centering
\parbox{\columnwidth}{\includegraphics[width=\columnwidth]{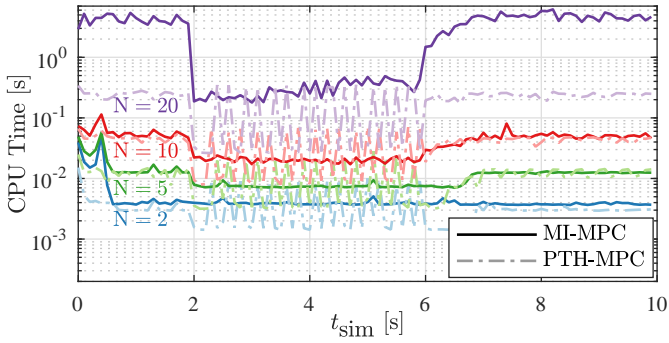}}
\caption{Computation (CPU) time for the trajectory in Fig. \ref{fig:ctrlTrajectory}a). The solid and dashed lines indicate the MI-MPC and the PTH-MPC controllers, respectively. The $N$ labels identify the two lines directly above them. Simulations were performed on an Intel(R) Core\textsuperscript{TM} i5 processor running at 2.40 GHz with 15.7 GB of useable RAM.}
\label{fig:compTime}
\end{figure}

\begin{table}[t!]
\caption{Maximum computation Time}
\centering
\begin{tabular}{cccccc}
\toprule
 & $N$ & 2 & 5 & 10 & 20 \\
\midrule
\multirow{2}{*}{CPU time [s]} & MI-MPC  & 0.051 & 0.057 & 0.114 & 6.171 \\
& PTH-MPC & 0.015 & 0.028 & 0.071 & 0.351 \\ 
\bottomrule 
\end{tabular}
\label{tab:compTime}
\end{table}

Across $t_{sim}$, the MI-MPC computation time remains relatively constant for all $x^r(t)$. However, when tracking an unreachable state reference, the PTH-MPC algorithm computation times fluctuate widely. This is due to Algorithm 1 requiring several iterations for the logarithmic barrier term to drive the calculated $u^*_{k|t}$ towards binary values. Increasing  $\gamma$ and $\beta$ reduces computation time, but assessing the impact this has on controller performance requires further study.

\section{Conclusion}
\label{sec:conclusion}
In this investigation, we demonstrated that MI-MPC and PTH-MPC are promising candidates for core fueling and density control in fusion tokamaks. Future work will evaluate and improve MPC prediction model accuracy against an experimentally validated transport model \cite{Blanken2018, Bosman2021}. For pellet injection control in ITER, the control decision set will need to be extended to include multiple launchers, variable fuel composition, and actuation uncertainty. We will also investigate further methods to increase controller efficiency to cope with the extended combinatorial optimization problem. 

\bibliography{LCSS_CDC_2023_Orrico}
\bibliographystyle{IEEEtran}

\end{document}